# Chiral light by symmetric optical antennas

Addis Mekonnen[1]\*, Tavakol Pakizeh[2], Irina Zubritskaya[1], Gustav Edman Jönsson[1], and Alexandre Dmitriev[1]\*

[1] Department of Applied Physics, Chalmers University of Technology, 412 96 Göteborg, Sweden

[2] Faculty of Electrical Engineering, K. N. Toosi University of Technology, Tehran 1631714191, Iran

\* Correspondence and requests for materials should be addressed to A. D., alexd@chalmers.se

and A. M., addis@chalmers.se

**Chirality is at the origin of life and is ubiquitous in nature. An object or a system is deemed chiral if it is non-superimposable with its own mirror image, which then relates to a particular way a circularly polarized light interacts with such object – something known as circular dichroism (CD), the differential absorption of right- and left-circularly polarized light[1,2]. According to the common understanding in biology, chemistry and physics, the chiroptical response results from an internal chiral structure or, in a special two-dimensional case, the extrinsic symmetry breaking under asymmetric illumination. Here we show that CD is possible with simple symmetric optical nanoantennas at symmetric illumination, that is – at normal incidence of light. We demonstrate that two electromagnetic dipole-like modes with a phase lag, in principle, suffice to produce far-field chiroptical response in achiral structure. This is experimentally and theoretically exemplified with the all-visible spectrum archetypical optical nanoantennas – symmetric nanoellipses and nanodimers, forming large-scale metasurfaces. The simplicity and generality of this finding reveal a whole new significance of the electromagnetic design and complex media engineering at a**

**nanoscale with potential far-reaching implications for optics in biology, chemistry and materials science.**

The conventional understanding of chiral optical activity such as CD and optical rotatory dispersion (ORD) in stereochemistry and life science is firmly linked to enantiomorphism, i.e., an object should be structurally non-superimposable by rotation and translation with its mirror image to display chiroptical effects. This fundamental concept from chemistry has been recently successfully adopted in nano-optics, where it aims at electromagnetic amplification of chiral sensing and novel nanophotonics devices based on modulation of polarization of light. Here plasmon metamolecules, engineered in two or three dimensions, form the basic electromagnetic modes of the nanoplasmonic elements to display high degree of CD [3-12]. In a special case a structurally achiral geometry can produce chiroptical response by asymmetric, typically off-normal illumination [13,14], structured illumination with the light carrying orbital angular momentum [15] or by forming a chiral superstructure [16]. Progressively more effort is focused on the optical near-field, where 'superchiral' enhanced near-fields are possible [17,18].

To outline in simple terms the commonly accepted concept of chirality in plasmonics and how the resulting CD links to the chiral structure, we consider one of the most basic plasmon chiral metamolecule: two coupled dipoles V-shaped nanoantenna [19,20] and its simplified orthogonal case – L-shaped nanoantenna (Figure 1a). It has been extensively employed for building functional metasurfaces [20,21] and is a prototypical geometry to visualize the principles of chiroptics with plasmonics. In practice, two coupled harmonic oscillators that represent basic dipolar modes of a plasmonic metamolecule, rooting in the Drude-Lorentz model for optical materials, is a very intuitive way to classically grasp the emergence of chiroptics and many other phenomena in nanoplasmonics and nano-optics [22-24]. The L-shaped nanoantenna can then be simply pictured as two masses on two springs that are connected via a third spring (Figure 1a top, also known as the Born-Kuhn model [23]). This results in a system that performs the rotation of a linearly polarized light by coupling it to the two orthogonal and, in general, identical dipolar oscillators with a phase lag, which emerges due to conductive or inductive coupling of the two. Naturally, the chiral shape of a nanoantenna, embedded in this mechanistic visualization, allows interacting differently with right- and left-circularly polarized light to produce CD (Figure 1a, bottom schematics).

In what follows we demonstrate that the phase difference between the two oscillators and thus the phase difference between the two dipole-like plasmon modes of a nanoantenna in itself is, in principle, the sufficient condition to earn chiroptical response from such optical nanoantenna (Figure 1b). Moreover, the distinct nanoantenna elements (Figure 1a, top) that present such plasmon modes are excessive, as a simple symmetric element can accommodate two spectrally distinct dipole-like (in this specific case - orthogonal) modes with intrinsic phase lag to display a marked chiral optical activity. Whereas a chiral plasmon nanoantenna affords the emergence of CD by the asymmetric spatial arrangement of nanoantenna elements (Figure 1a, bottom), the extraordinary chiroptical response can be earned simply with two spectrally separated dipole-like plasmon modes (Figure 1b, bottom) in a symmetric nanoantenna. Figure 1c then pictures the fabricated exemplary symmetrical plasmon elliptical nanoantennas, positioned in a short-range ordered array and forming a macroscopic metasurface of non-interacting individual elements, set to display extraordinary chiral optical response.

We produce macroscopic short-range ordered arrays of elliptical plasmon nanoantennas on glass slides via bottom-up nanofabrication with hole-mask colloidal lithography (Figure 1c) [25]. Elliptical optical nanoantennas generate a typical size-dependent optical response at two orthogonal polarizations, corresponding to light coupling to two dipole modes – longitudinal and transversal, i.e., along the long and the short axes of the nanoellipse, respectively. Left panels of Figures 2a-c depict a sequence of size-dependent optical extinction (green and red spectra) along the two orthogonal axes of the nanoantennas of different lateral dimensions and the same thickness of 20 nm (nanoantennas schematics and lateral dimensions are also shown on the left side of each figure). The optical response of the similarly sized nanodisk antenna is also included in Figure 2e (see also SEM inset for the structure of nanodisks array, which is similar to that of nanoellipses). The key feature of a basic chiral L-nanoantenna is the presence of two coupled dipolar modes in orthogonal non-symmetric configuration. Here with a symmetric elliptical patch nanoantenna we test whether simply two oscillators (in general, not coupled) suffice to create a chiroptical respose. The blue data points in the left panels of Figures 2a-c feature a sequence of CD spectra, collected on these nanoellipse optical antennas arrays with commercial CD spectrometer (see Suppl. Info). That is, similarly to the conventional biochemical CD measurements, the right- (RCP) or left- (LCP) circularly polarized light is incident normal ontoly the nanoantennas metasurface and the change in transmission $\Delta T$ of the two ($\Delta T = T_{RCP} - T_{LCP}$) is expressed in mdeg. We should note here that typically HCL-produced short-range ordered arrays display no presence of near-field coupling and only very weak far-field interaction via scattering [26], making such arrays essentially spectrally representative of the constituent individual nanoantennas. Strikingly, not only is the CD here non-zero (thus coined 'extraordinary'), but it is also clearly size- and resonance-dependent, reaching rather significant values ($\approx 0.1$ deg.) for a certain nanoantenna size (see 100 nm x 140 nm configuration in Fig. 2c). Here with conventional circular-polarized light transmission measurements we produce chiroptical response with symmetric nanoantennas at symmetric illumination. Note that in the case of a nanodisk antenna the CD is not detected (Figure 2e) as two potentially orthogonal dipolar modes of a nanodisk are identical in phase and amplitude. We compliment the detection of the CD with symmetric nanoantennas by collecting the ORD by using linearly polarized light at normal incidence (Fig. S1, Suppl. Info). Indeed, as both phenomena are closely related, we straightforwardly obtain a sequence of the ORD spectra for the considered nanoelliptical and nanodisk antennas. We further stay with ORD and probe its dependence on the angle of the linear polarization of the incoming light with respect to nanoellipse antenna axes (Fig. 2S, Suppl. Info). While two orthogonal dipolar modes of the nanoantenna are not coupled (as schematized in Fig. 1b), the ORD goes to zero when linearly polarized light is aligned with one of the two nanoantenna axes, and reaches its maximum roughly at 45 deg. orientation, when both dipolar modes of the nanoellipse are engaged (Fig. 2S).

To rule out any geometrical asymmetry of nanoelliptical antennas due to fabrication or the presence of effective chiral superstructure in nanoantennas array, we check for the reciprocity of the generated CD – i.e., for the planar chiral structure inverting the incident angle of light (from 0 deg. to 180 deg.) would necessarily result in CD inversion as the structure adopts its mirror image. However, no CD inversion is detected for the present nanoelliptical antennas (see Fig. 3S, Suppl. Info). This makes these nanoelliptical optical nanoantennas essentially 'truly 3D chiral'.

To establish a simple framework for the emergence of extraordinary chiroptics in symmetric electromagnetic systems we turn again to the mechanical analogue of the dipolar plasmon modes in a nanoantenna as a pair of orthogonal oscillators (see cartoon schematics in Fig. 1b and Fig. 3S, Suppl. Info). By contrast with L-nanoantenna (Figure 1a), in a symmetric elliptical patch nanoantenna the two orthogonal uncoupled oscillators are acting within the same element. Polarizing one oscillator stronger than the other would result in a displacement of the effective 'center of mass' of the system (see Fig. 4S, Suppl. Info – the displacement is realized by introducing a certain displacement angle $\Theta$). Note that the effective center of mass emerges only when both oscillators are engaged (see Fig. 4S). What would that mean for the chiral optical response? We plot the two basic orthogonal dipolar electromagnetic modes of the patch nanoantenna (longitudinal and transversal) using Drude-Lorentz formalism (right panels of Figs. 2a-c and Fig. 2e), which is conventionally employed to model the optical response of nanoplasmonic systems in the realm of classic electrodynamics (see Methods Summary and Suppl. Info). We note that we obtain reliable qualitative fit between the calculated optical extinction and experimental data on nanoelliptical and nanodisks antennas (compare red and green on the left (experimental data) and right (simulated data) panels of Figs. 2a-c,e). The CD that the system produces (blue simulated data of Figs. 2a-c and e, displayed as phase, in mdeg) is then plotted using the polarizability of the two dipolar oscillators considering the displacement of the effective center of mass. In a simple mechanical description we can picture two oscillators driven by different external forces, which in practice would mean that the two dipolar modes of a nanoantenna have different in-coupling strengths (extinction efficiency) for the incoming light at particular wavelength. Now imagine driving the system simultaneously along $x$ and $y$ axes, but mainly spectrally matching the 'red' oscillator resonance (Fig. 1b and Fig.4S bottom, Suppl. Info). In practice this is realized either with linearly polarized light off the axes of the nanoantenna, or straightforwardly with circularly polarized light. This excitation makes the 'red' oscillator to displace more along the *x*-axis than the "green" one does along *y* since the former has higher efficiency at this chosen wavelength. This causes the effective center of mass in turn to displace. In electrodynamics terms, we gain larger polarizability $\alpha_1$ of the 'red' oscillator, and smaller polarizability of the green one $\alpha_2(\omega)$. Back to the mechanistic model, the displacement angle of the center of mass from the *x*-axis can be calculated by $\theta_2 \approx tan^{-1}\left(\frac{\xi\alpha_2(\omega)}{\alpha_1(\omega)}\right)$ with respect to the incident polarizing (driving) field along *x*, with $\xi$ as a fitting parameter.

As a consequence, if a dipole oscillates at an angle to an incident driving electromagnetic field, the resulting field gets rotated and obtains a phase difference with the incoming one. In this way, we arrived at two defining features of all chiroplasmonic nanoantennas: that is, they provide polarization rotation, also interpreted as ellipticity, accompanied by the modification of the phase of the incident electromagnetic wave. Note that, in general, in this line of reasoning the mutual orientation of two oscillators in terms of asymmetry is not considered at all. Similarly to the outlined scenario for the 'green' oscillator, the 'red' one (Fig. 1b and Fig. 4S top, Suppl. Info) would experience the same tilt at the 'green' driving frequency and will be rotated out of the *y*-axis by the angle $\theta_1 \approx tan^{-1}\left(\frac{\xi\alpha_1(\omega)}{\alpha_2(\omega)}\right)$. Over the whole frequency range of interest, the effective tilt angle of the resulting electromagnetic field would emerge $\theta_{eff} = \theta_1 + \theta_2 = tan^{-1}\left(\frac{\xi\alpha_1(\omega)}{\alpha_2(\omega)}\right) + tan^{-1}\left(\frac{\xi\alpha_2(\omega)}{\alpha_1(\omega)}\right)$, summing up the contributions of both dipolar modes of the nanoantenna. Here $\theta_{eff}$ is a complex angle with real $\Re(\theta_{eff})$, and imaginary $\Im(\theta_{eff})$ parts that represent the ORD and CD, respectively. Using a fitting parameter (set to $\xi = 0.01$), the CD response is modeled for each nanoellipse and nanodisk antennas and plotted in mdeg in Figs. 2a-c and e (blue data points). Overall we clearly reproduce experimental data features with this simple two-oscillator model (compare left and right panels of Figs. 2a-c, e).

As the phenomenon of chiral light generation by symmetric optical antennas relies purely on the phase difference between the two oscillating electromagnetic subsystems and not on their structural details, we aim to test it with more complex nanoantennas designs. The plasmonic nanodisks dimer is a well-studied nanoantenna type, where a mode hybridization scenario between the basic dipolar plasmon modes of individual nanodisks in the pair is realized [27,28]. A needed pair of the resulting hybrid modes that would deliver the needed phase lag to produce chiroptical response of this optical antenna will be those along and perpendicular the dimer axis. We experimentally consider a general case of a dimer nanoantenna made of two different (magneto)plasmonic materials Ni and Co [29], each sized 150 nm in diameter and 30 nm in height, with the gap size of approximately 10 nm (see SEM inset of Figure 2d). The extinction spectra along (red) and perpendicular (green) the dimer antenna axis display broad plasmonic features (Fig. 2d, left panel), characteristic for plasmonic nanoferromagnets [29]. As with nanoelliptical patch antenna, here the CD emerges in the spectral range of the hybrid nanodimer resonances (left panel of Fig. 2d, blue data points). Further, this CD by nanodimer antennas can be reproduced with our two-oscillator model with the fitting parameter $\xi = 0.001$ (Fig. 2d, right panel).

It is known that breaking the space inversion symmetry might be promoted by the radiation of an electromagnetic dipole near an interface. All studied here symmetric nanoantennas are produced on a glass substrate, so it is natural to verify the potential role of air-glass interface on the emergence of chiroptical response from such antennas. For this we embed nanoantennas into a transparent thermoplastic polymethyl methacrylate (PMMA) that has a refractive index of 1.5 largely matching that of the glass substrate in the visible spectral range. The resulting CD is shifted to longer wavelength (as expected), and overall has even doubled its intensity (Fig. 5S, Suppl. Info).

It is generally accepted that CD in electromagnetic systems emerges when both space and time inversion symmetry are broken along the light propagation direction. The presented results with symmetric optical antennas under symmetric illumination essentially suggest that just breaking the time inversion symmetry suffices to obtain CD, with the phase lag acting as symmetry-breaking agent. In this line of thought it is the 'phase (or time) asymmetry' that we probe here, which, nevertheless, evidences itself just like the space asymmetry does – that is, by bestowing a chiroptical response to light. As a matter of fact, some analogy to this could be found in magneto-optical systems, where the application of the magnetic field breaks the time reversal symmetry in otherwise structurally symmetric materials and produces non-reciprocal propagation of light. Further, in these materials the phenomenon of magnetic circular dichroism (MCD) is very well known [30]. That is, the CD is gained in structurally symmetric materials by inducing the magnetic polarization. Here as we are attempting to conceptualize the whole new importance of breaking the time-reversal symmetry in electromagnetic systems to produce chiroptical response, by simple mathematical transformation with physical interpretation we could try to link the actual CD with the phase lag. Indeed, if one considers our calculated extinction of the two 'red' and 'green' dipolar modes of the nanoantenna (right panel of Fig. 2c and Fig. 6Sa, Suppl.

Info), their phases (again, 'red' and 'green') and the corresponding phase difference ('blue') can be separately plotted (Fig. 6Sb, Suppl. Info). Note that the CD crosses zero when the phase difference reaches maximum for this set of dipolar modes at 90 deg. ($\pi/2$). Interestingly, taking a simple derivative of this phase difference with respect to the wavelength we arrive at something looking qualitatively similar to what we plotted earlier as the CD (compare 'blue' curves in Figs. 6Sa and 6Sc, Suppl. Info). As the derivative is the instantaneous rate change, one could even tentatively establish a phenomenological link between the *rate change of the phase lag in wavelength* with the spectrally-resolved CD.

To further support the experimentally detected emergence of the CD in symmetric electromagnetic antennas, we perform the numerical calculations by means of the full-wave three-dimensional dispersive finite-difference time-domain (FDTD) method based on the rigorous Maxwell equations (further details are found in Methods Summary and Suppl. Info). We obtained the resonant dipolar modes of nanodisk (Fig. 3a) and nanoellipse (Fig. 3b) antennas, illuminated by an optical Gaussian pulse with circular polarization and broad spectral coverage in Fourier domain. The optical CD, according to its most general definition, is then obtained as a difference of the spectral optical extinction for the RCP and LCP light, i.e. $\text{CD} = Q_{ext}^{\text{RCP}} - Q_{ext}^{\text{LCP}}$ (Fig. 3c). From the latter we see that whereas the calculated CD from the nanodisk antenna (dia. 100 nm, thickness 20 nm) is basically absent ($\sim 10^{-15}$), the CD from a nanoellipse antenna (140 nm x 100 nm, thickness 20 nm), raises up to a detectable level ($\sim 10^{-4}$), thus confirming also numerically the observation of the chiral light generation by a completely geometrically symmetric electromagnetic system.

Finally, we tackle the physical meaning of the tilt angles of the mechanical oscillators in our two-oscillator model. Though the presented reasoning is rather simplified, it is intriguing to find an analogy of the envisioned tilt angles in real and measurable physical values. According to the model, $\theta_1$ and $\theta_2$ present a tilt to opposite directions. If we picture the incoming light at a normal incidence and linearly polarized to form some angle with both dipolar modes of a nanoellipse antenna (Figure 1b) to ensure their simultaneous excitation, the outcoming light in the forward direction would be polarized at a certain angle $\theta_{eff}$, which would combine those $\theta_1$ and $\theta_2$. The generated CD is, again, proportional to $\Im(\theta_{eff})$. The presence of the two distinct angles for the polarization rotation ($\theta_1$ and $\theta_2$) that dominate two distinct excitation wavelengths ('green' and 'red') in practice would mean that the polarization rotation becomes color-selective.

To probe this experimentally we collect the transmission of a linearly polarized light using a polarizer-analyzer combination (Figure 4a). If we assume from the model that the linearly polarized light gets rotated by the nanoellipse antennas array and this rotation is, in addition, spectrally selective, then by adjusting the analyzer we would detect distinctly colored and polarized light after it passed the array (Figure 4a – see also the photograph of the actual used sample and its SEM image of the 100 nm 140 nm nanoellipse antennas array). First, we position the analyzer to earn fully extinct light (i.e., at 90 deg. to the polarizer) on the other side of the clean glass slide (Figure 4b, black data points). When introducing the glass slide with nanoellipse antennas (their longitudinal axis at 45 deg. to the incoming linear polarization) we produce a marked increase in the transmitted intensity (Figure 4b, blue data points) in the spectral region where two dipolar modes of the nanoantenna are active (c.f., Figure 2c), which is essentially the manifestation of ORD. This means that the efficient rotation of the incoming polarization is performed. Next, we position the analyser at a certain angle to the polarizer ($\theta_a$ on the schematics, here 1 degree – that is, the light is not fully extinct) and collect the transmitted baseline from the glass slide (Figure 4c, black data). When nanoellipse antennas are introduced, the transmission is dramatically modified (Figure 4c, turquoise and magenta data points). For example, at around 720 nm (i.e., close to the 'red' nanoellipse resonance as in Figure 2c) only a particular polarization rotation angle produces the most of the transmitted light (magenta), whereas for polarization at all others the transmission is suppressed. Rotating the analyzer to an opposite angle $-\theta_a$ allows us to favor the transmission for the opposite polarization rotation angle (turqiose) at another chosen wavelength (610 nm) close to the 'green' dipolar mode (see Figure 2c). So the linear polarization of the incoming wave with a certain wavelength, close to one of the dipolar modes, would be rotated to a certain angle, defined by the interplay of the polarizabilities of the two dipolar modes of the nanoantenna and their coupling. Essentially, the constructed conceptual device functions as a 'color-selective polarizer' that at each given wavelength produces given polarization rotation angle.

Summarizing, we demonstrate how the prerequisite of a chiral geometry is lifted for optical nanoantennas to produce chiroptical response. We show both experimentally and theoretically how a symmetric single-element nanoelliptical antenna displays sizeable CD at normal illumination. This makes us to conclude that the condition for the system of two dipolar modes to produce the optical activity is the phase difference between the modes. We verify this idea with a simple two-oscillators model and numerical simulations. Further, we bring forward a concept of the color-selective polarizer that enables spectrally resolved polarization rotation of the incoming linearly polarized white light. For the latter, our bottom-up nanofabrication delivers truly large-scale $cm^2$ surfaces of symmetric elliptical nanoantennas that are readily available for further integration to the potential optical devices. Apart of intriguing consequences in the fundamental understanding of the chiral interaction of light with various electromagnetic systems, here a simple symmetric nanogeometry with high chiroptical activity opens unforeseen opportunities in chirophotonics to design flat optical elements for light management, electromagnetic engineering of complex media and in chemical and biological chiral sensing.

**Acknowledgment**

A. M. acknowledges Knut and Alice Wallenberg Foundation; G. E. J. acknowledges Swedish Foundation for Strategic Research (SSF) programme RMA011 'Functional electromagnetic metamaterials for sensing'; I. Z., K. L. and A. D. acknowledge SSF Future Research Leader grant.

**Author contribution**

A. M. and A. D. devised the concept; A. M. and I. Z. performed the optical measurements; I. Z., G. E. J. and A. M. fabricated the samples; T. P. performed FDTD calculations; A. M. developed the analytical model; A. M. and A. D. wrote the manuscript; all authors contributed to the discussions.

**Competing financial interests**

The authors declare no competing financial interests.

Figure 1

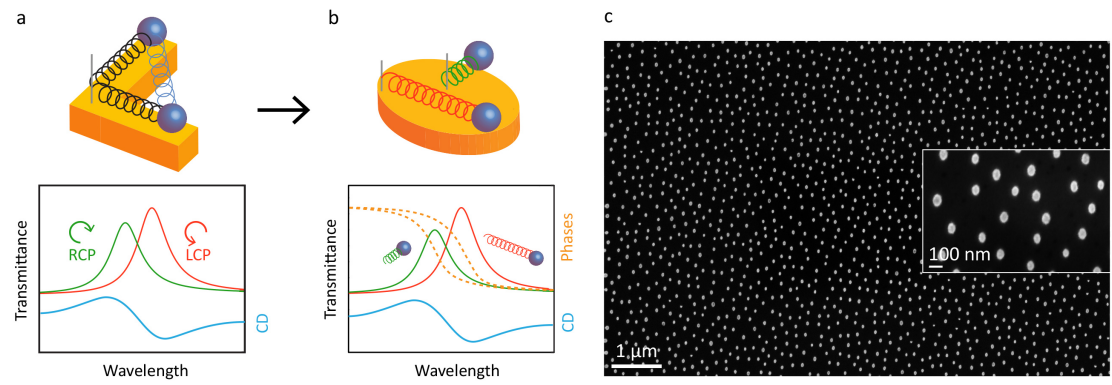

Figure 2

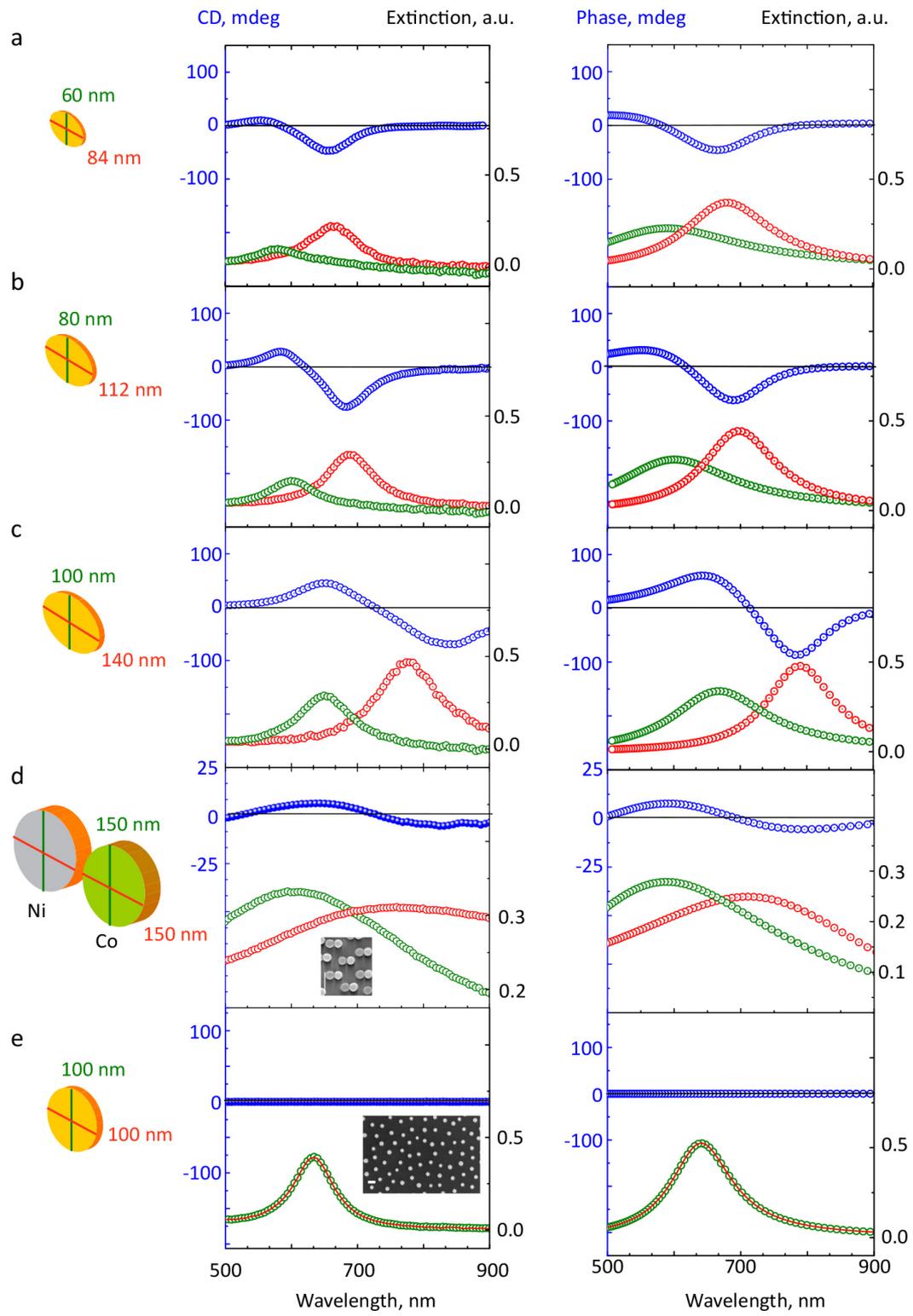

Figure 3

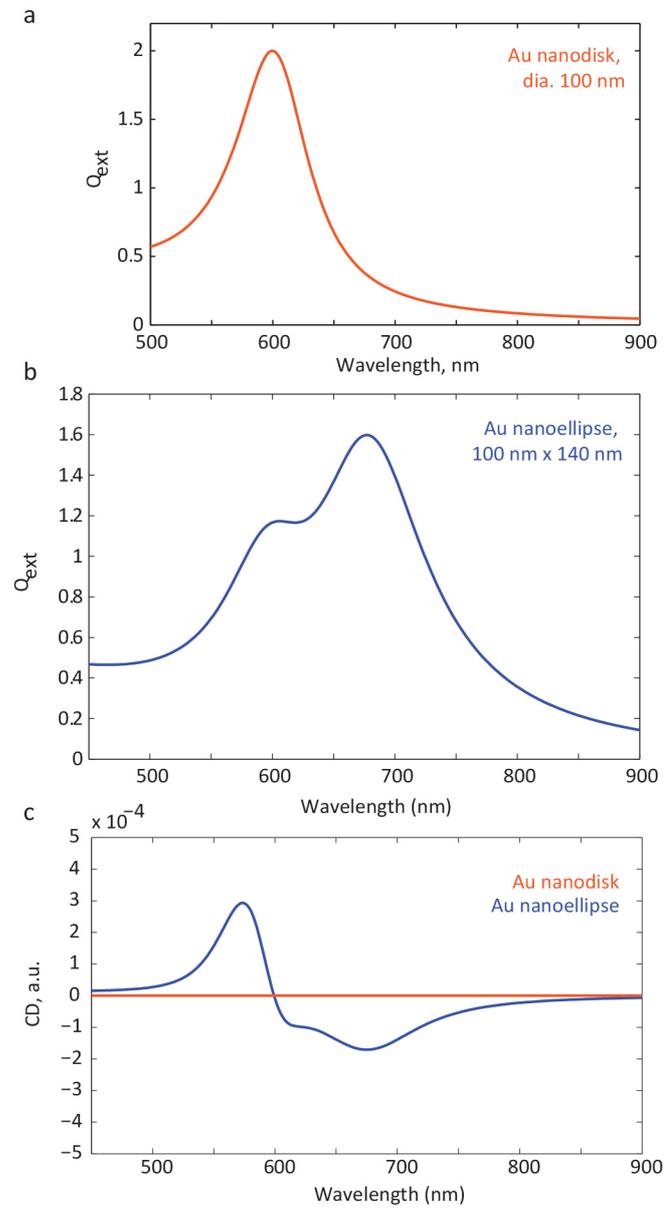

Figure 4

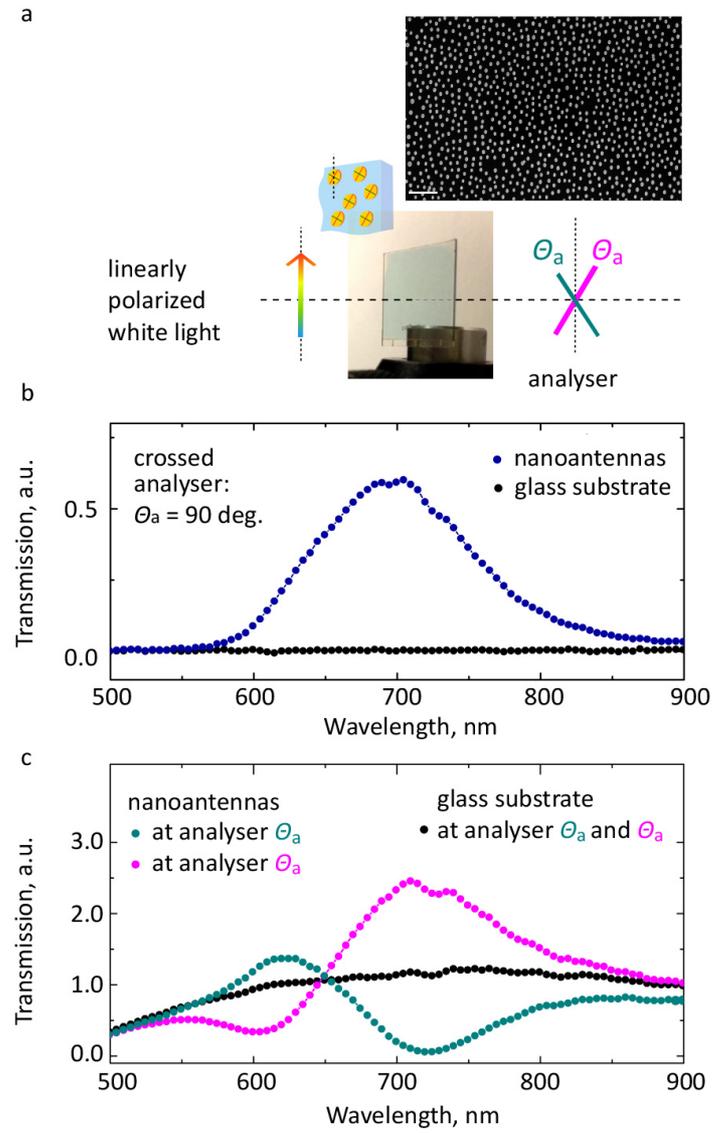

Figures captions

**Figure 1: The prerequisite of a chiral structure to produce chiroptical response is removed in the simplest case of two-dipole-modes optical nanoantennas.** (a) Archetypical planar chiral L-shape nanoantenna, effectively represented by two coupled dipolar oscillators (top), produces chiroptical response (CD) when illuminated with right- (green) and left- (red) circularly polarized light (bottom). (b) The chiroptical response can be earned instead by using two spectrally distinct oscillators (green and red spectra in the bottom, usually obtained with linearly polarized light) in a completely symmetric configuration (nanoellipse optical antenna, top schematics) due to their in-built phase difference (corresponding phases of the two oscillators are shown on the bottom schematic graph). (c) Scanning electron micrograph, overviewing a typical macroscopic short-range-ordered arrangement of symmetric nanoelliptical optical antennas (see the inset for a closer look at the shape of the individual nanoantennas), employed throughout this work.

**Figure 2: Chiroptics with symmetric nanoantennas.** Left panels: (a-c, e) Experimental optical extinction (right axis, green and red data) and CD (left axis, blue data) at normal light incidence of the basic plasmon dipolar modes (marked green and red with dashed lines) of nanoelliptical and nanodisk antennas (see corresponding geometries and sizes on cartoons to the left). Inset in (e) - SEM of nanodisks (scale bar 200 nm) short-range-ordered bottom-up array. Right panels: (a-c, e) Corresponding modeled optical extinction (right axis) and the CD (left axis) for two separate dipolar modes of nanoellipses and nanodisks. d) Experimental extinction of the Ni-Co dimer antennas (blue data, left axis) (left panel). Inset – SEM (scale bar 400 nm) of the dimer nanoantennas (cartoon schematic to the left). Modeled optical

response of individual Ni-Co dimer (right axis) and its corresponding CD (left axis) (right panel).

**Figure 3: FDTD-simulated generation of the chiral light from symmetric nanoantennas.** Extinction efficiency of the nanodisk optical antenna (a) and nanoellipse antennas (b) upon illumination with circularly-polarized light. When comparing the illumination with LCP and RCP, nanodisk antenna display the same extinction efficiency for both, whereas nanoellipse shows the generation of pronounced CD in the visible spectral range (c).

**Figure 4: A color-selective polarizer.** (a) The concept of a color-selective polarizer. Surface area of the optical element (photo) – 1.5 x 1.5 cm$^2$. SEM scale bar – 1 um (b) Transmission in a cross-polarized geometry ($\theta_a$ = 90 deg.) on bare glass slide (black) and nanoellipses array (blue). (c) Analyzer rotation to +/- 1 deg. allows to detect one (magenta) or another (turquoise) polarization rotation at a particular wavelength as transmission. With rotated polarizer clean glass slide displays a certain transmission that we take as a reference (black).